\documentclass[aps,pre,reprint,nofootinbib,superscriptaddress,floatfix]{revtex4-2}

\usepackage{amsmath,amssymb,bm}
\usepackage{mathtools}
\usepackage{booktabs}
\usepackage{graphicx}
\usepackage[colorlinks=true,linkcolor=blue,citecolor=blue,urlcolor=blue]{hyperref}

\newcommand{\Uext}{\mathcal U_{\rm ext}}
\newcommand{\Ucond}{\mathcal U_{\rm cond}}
\newcommand{\Agap}{\mathcal A}
\newcommand{\Cgap}{\mathcal C}

\begin{document}

\title{Path-Extrema Upper Bounds on Mean Entropy Production}

\author{Surachate Limkumnerd}
\affiliation{Department of Physics, Faculty of Science, Chulalongkorn University, Bangkok 10330, Thailand}

\date{\today}

\begin{abstract}
Fluctuation relations imply $\langle\Sigma_T\rangle\ge0$. What can path extrema tell us about an upper bound? A ceiling on the positive entropy-production maximum alone gives no improvement over the trivial endpoint bound: rare negative excursions can supply the exponential weight required by the fluctuation relation. Given the joint law of the running extrema $L_T$ and $H_T$ of $M_t=e^{-\Sigma_t}$, we derive an upper bound $\Uext$ by allocating a normalization budget across realized intervals, favoring those that yield the most entropy per unit cost. The exact identity $\langle\Sigma_T\rangle=\Uext-\Agap-\Cgap$ separates an allocation gap across extrema classes from a curvature gap within each class. The bound and decomposition require the terminal relation $\langle M_T\rangle=1$ and the pathwise extrema constraint, not martingality; $\Uext$ is sharp in this minimal class. Martingality imposes further constraints, and attainment of the same optimum under them remains open in general. For a three-state Markov-jump cycle, we obtain the finite-time joint law of $(\Sigma_T,L_T,H_T)$ by reduction to a biased random walk. A companion Monte Carlo study illustrates how undersampling rare negative excursions prevents the fluctuation relation from being resolved at long times. The result is a structural upper constraint, not a more efficient estimator of the mean from complete trajectories.
\end{abstract}

\maketitle

\section{Introduction}

The integral fluctuation theorem $\langle e^{-\Sigma_T}\rangle=1$ implies the second law $\langle\Sigma_T\rangle\ge0$ \cite{Seifert2005}. Thermodynamic uncertainty relations, speed limits, and inference methods use additional information to bound entropy production from below \cite{BaratoSeifert2015,Gingrich2016,HorowitzGingrich2020,Shiraishi2018,DechantSasa2018,Bisker2017,Seifert2019,Harunari2022,Singh2024}. Here we ask the complementary question: how large can the mean be when only path extrema are retained? Previous upper bounds use entropy extrema and convexity \cite{Limkumnerd2017}, or kinetic information \cite{NishiyamaHasegawa2023}. Martingale stochastic thermodynamics provides a related perspective on extrema and stopping times \cite{Roldan2015,Neri2017,Pigolotti2017,Neri2019,Neri2020,Manzano2022,Roldan2024}.

Let $M_t=e^{-\Sigma_t}$, with running extrema $L_T=\inf_{t\le T}M_t$ and $H_T=\sup_{t\le T}M_t$. A ceiling on the positive entropy-production maximum alone gives no improvement over the trivial endpoint bound: rare negative excursions can supply the exponential weight needed for normalization. Retaining the joint law of both extrema yields a nontrivial bound $\Uext$, obtained by ranking intervals by entropy gain per normalization cost. The gap identity $\langle\Sigma_T\rangle=\Uext-\Agap-\Cgap$ separates suboptimal allocation across extrema classes from the curvature cost of terminal values inside their realized intervals.

The bound and decomposition require the terminal fluctuation relation and pathwise extrema constraints, not martingality. The bound is sharp in this minimal class, but its attainment under the stronger martingale constraint is a separate question. We illustrate the result with an exactly solved three-state cycle and use Monte Carlo sampling to demonstrate the difficulty of resolving rare events. Supporting derivations and examples are in the Supplemental Material \cite{SupplementalMaterial}.

\section{Setup: minimal hypothesis and the martingale specialization}
\label{sec:setup}

\subsection{Minimal hypothesis}
\label{sec:setup-minimal}

Let $\Sigma_t$ denote the stochastic entropy production accumulated up to time $t$ along an observed trajectory, and set $M_t=e^{-\Sigma_t}$. The static upper bound and the gap decomposition proved in Secs.~\ref{sec:random-envelope}--\ref{sec:decomposition} use only the following minimal hypothesis:
\begin{itemize}
\item[(H1)] $M_t>0$ for $t\in[0,T]$, with $M_0=1$;
\item[(H2)] the terminal relation $\langle M_T\rangle=\langle e^{-\Sigma_T}\rangle=1$, i.e., the fixed-time integral fluctuation theorem;
\item[(H3)] the running extrema $L_T=\inf_{0\le t\le T}M_t$ and $H_T=\sup_{0\le t\le T}M_t$ are recorded, with $\langle|\ln M_T|\rangle$, $\langle|\ln L_T|\rangle$, $\langle|\ln H_T|\rangle$, and $\langle H_T\rangle$ all finite.
\end{itemize}
Because $T\in[0,T]$, (H3) implies the pathwise sandwich $L_T\le M_T\le H_T$ automatically, for any process with recorded running extrema. No conditional expectation of $M_t$ given the past, and no property of $M_t$ at intermediate times, enters the derivation of $\Uext$ or of $\langle\Sigma_T\rangle=\Uext-\Agap-\Cgap$. Section~\ref{sec:discussion} summarizes which weaker or stronger statement each remaining result actually uses.

\subsection{The martingale specialization}
\label{sec:setup-martingale}

A stronger and physically central special case is that in which $M_t$ is a positive martingale with respect to the natural trajectory filtration, $\langle M_t\mid\mathcal F_s\rangle=M_s$ for $0\le s\le t\le T$. This holds for stationary Markov jump processes with local detailed balance and for diffusions under the usual stochastic-thermodynamic regularity conditions \cite{Seifert2012}, whenever $\Sigma_t$ is the entropy production relative to the steady-state dynamics: $\Sigma_t$ is then the log ratio between the probability of the realized trajectory and its time-reversed counterpart. Hence $e^{-\Sigma_t}$ is a Radon--Nikodym density process and automatically a positive martingale with $M_0=1$, from which (H2) follows as the case $t=T$. The martingale property is not needed for (H1)--(H3); its physical role is to supply (H2) automatically, by pinning the conditional future average of $M_t$ to its present value at every time, not merely at the terminal time.

The martingale property also places the results alongside recent universal results on entropy-production infima, stopping times, and extreme-value statistics \cite{Roldan2015,Neri2017,Pigolotti2017,Neri2019,Neri2020,Manzano2022,Roldan2024}. Sharpness of $\Uext$ within the unrestricted (H1)--(H3) class does not need it: the Supplemental Material constructs a process that attains $\Uext$ without martingale structure. Under martingale constraints, the deterministic two-sided support bound can be attained by stopping at two barriers (Sec.~\ref{sec:deterministic-two-sided}); attainment of the random-extrema bound for a prescribed extrema law remains a separate question (Sec.~\ref{sec:discussion}).

\subsection{Nonstationary and driven systems}
\label{sec:setup-nonstationary}

For explicitly time-dependent or driven systems, the static bound and decomposition apply verbatim under (H1)--(H3) whenever the chosen entropy-production functional $\Sigma_t$ obeys the fixed-time terminal relation $\langle e^{-\Sigma_T}\rangle=1$, whether or not $e^{-\Sigma_t}$ is a martingale at intermediate times. For driven Markovian systems, the usual total entropy production is defined through the log ratio of the forward trajectory measure to that of the appropriately time-reversed process, for which the fixed-time integral fluctuation theorem is standard \cite{Seifert2012}. Ref.~\cite{Roldan2024} discusses the more delicate nonstationary martingale structure. Only the martingale-specific content of Sec.~\ref{sec:setup-martingale} -- the automatic supply of (H2), the stopping-time constructions, and the connection to the extrema/stopping-time literature -- can fail away from stationarity.

\subsection{Running extrema}
\label{sec:setup-extrema}

We define the running lower and upper extrema of $M_t$ by
\[
L_T=\inf_{0\le t\le T}M_t,
\qquad
H_T=\sup_{0\le t\le T}M_t .
\]
Since $M_0=1$, we have $L_T\le 1\le H_T$ almost surely. The corresponding entropy-production extrema are
\[
A_T=-\ln L_T=\sup_{0\le t\le T}\Sigma_t,
\quad
B_T=\ln H_T=-\inf_{0\le t\le T}\Sigma_t .
\]
Thus $A_T$ is the largest positive entropy production reached by the trajectory and $B_T$ is the magnitude of the largest negative entropy production reached by the trajectory.

\begin{figure}
\includegraphics[width=\columnwidth]{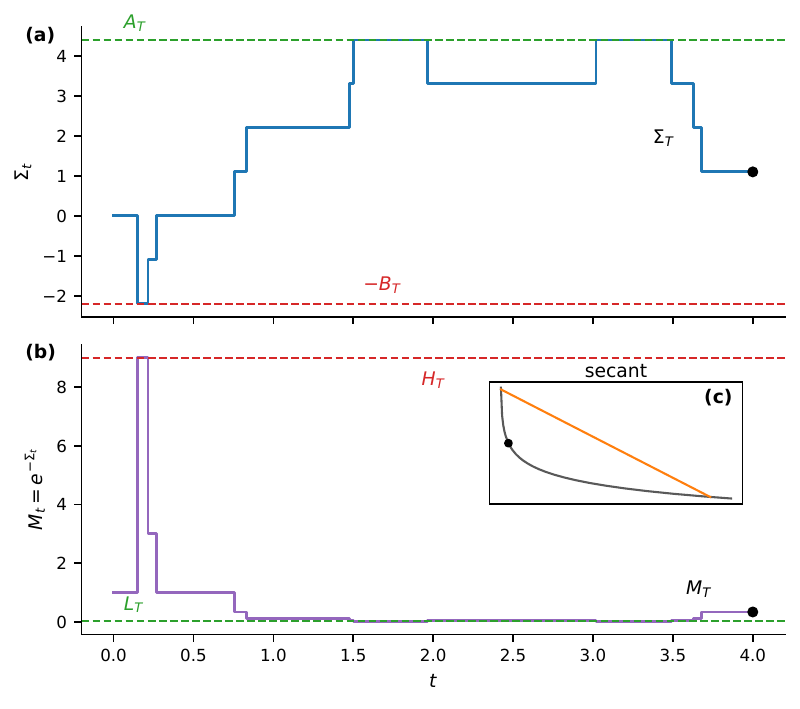}
\caption{Trajectory geometry. Panel (a) shows a sample entropy-production trajectory $\Sigma_t$, with the largest positive excursion $A_T$ and the largest negative excursion $-B_T$. Panel (b) shows the corresponding exponential process $M_t=e^{-\Sigma_t}$, whose running minimum and maximum define the realized extrema interval $[L_T,H_T]$; the terminal value $M_T$ lies inside this interval. Panel (c) shows the local convexity geometry: on a fixed realized interval, the convex function $-\ln m$ is bounded above by the secant joining $(L_T,-\ln L_T)$ and $(H_T,-\ln H_T)$, which represents the conditional terminal mean by an effective endpoint mixture.}
\label{fig:geometry}
\end{figure}

Figure~\ref{fig:geometry} illustrates these three pieces: the entropy-production extrema in panel (a), the extrema interval in panel (b), and the secant interpretation of the convexity bound in panel (c). The distinction between terminal support and path extrema matters: a deterministic support condition such as $e^{-a}\le M_T\le e^b$ constrains only the final value, while the random pair $(L_T,H_T)$ records the actual lower and upper extrema reached along each path. The latter contains strictly more path information, but also obeys additional constraints because the interval is created by the path itself. (H3) restricts attention to paths with $0<L_T\le M_T\le H_T<\infty$ and finite logarithmic moments; these can be relaxed by standard truncation arguments.

On the event $L_T=H_T$ the process is constant over $[0,T]$ and $M_T=L_T=H_T$; such paths contribute no interval width, and whenever a ratio involving $H_T-L_T$ appears below it is understood to be evaluated on $L_T<H_T$, with zero contribution assigned on $L_T=H_T$.

For each realized interval, replace the conditional terminal mean by an equivalent mixture of its two endpoints. The fluctuation relation will then fix the total normalization cost available for allocation. Define
\[
\mu_T=\mathbb E[M_T\mid L_T,H_T] .
\]
Since $L_T\le M_T\le H_T$, it follows that $L_T\le \mu_T\le H_T$.
We then define the extrema-conditioned endpoint weight
\begin{equation}
r_T=\frac{H_T-\mu_T}{H_T-L_T}.
\label{eq:r_actual}
\end{equation}
This quantity satisfies $0\le r_T\le 1$ and is the unique number for which $\mu_T=r_T L_T+(1-r_T)H_T$.
Thus $r_T$ is the effective conditional weight that would reproduce the same conditional terminal mean by placing mass only at the two endpoints $L_T$ and $H_T$.

The terminal relation (H2) imposes a global constraint on $r_T$: taking the expectation of the previous relation gives
\[
1=\langle M_T\rangle=\langle \mu_T\rangle
=
\langle r_T L_T+(1-r_T)H_T\rangle ,
\]
using only the tower property of conditional expectation, so
\begin{equation}
\langle r_T(H_T-L_T)\rangle=\langle H_T\rangle-1 .
\label{eq:rt_constraint}
\end{equation}
This identity is the constraint behind the path-extrema upper bound. It expresses the amount by which terminal weight must be shifted from the upper endpoint $H_T$ toward the lower endpoint $L_T$ in order to satisfy $\langle M_T\rangle=1$; it uses (H2) and the tower property alone, not the full martingale property.

The entropy-production interpretation of $r_T$ is direct. The endpoint $L_T=e^{-A_T}$ corresponds to the positive entropy-production extreme $\Sigma=A_T$, while $H_T=e^{B_T}$ corresponds to the negative entropy-production extreme $\Sigma=-B_T$. A larger value of $r_T$ means that, conditional on the realized interval, the terminal value is biased more strongly toward the entropy-producing endpoint.

We will also use the secant line of the convex function $f(m)=-\ln m$ over the interval $[L_T,H_T]$:
\[
\ell_{L_T,H_T}(m)
=
\frac{H_T-m}{H_T-L_T}(-\ln L_T)
+
\frac{m-L_T}{H_T-L_T}(-\ln H_T).
\]
For $m\in[L_T,H_T]$, convexity gives $-\ln m\le \ell_{L_T,H_T}(m)$.
This elementary inequality is the local building block of all the upper bounds below. The Supplemental Material collects the notation introduced in this section.

\section{Positive extrema are not enough: deterministic two-sided bound}
\label{sec:deterministic}

Before treating random path extrema, it is useful to separate two simpler facts. The first shows that positive entropy-production extrema alone cannot give a nontrivial upper bound. The second is the deterministic two-sided bound obtained when both positive and negative entropy-production ranges are controlled.

\subsection{The one-sided ceiling}

Suppose that the positive entropy-production maximum is bounded by a deterministic value $a>0$: $A_T=\sup_{0\le t\le T}\Sigma_t\le a$. Then, trivially, $\Sigma_T\le A_T\le a$, and hence $\langle \Sigma_T\rangle\le a$.
The important point is that no universal improvement below $a$ is possible.

To see this, let $\ell=e^{-a}$ and consider, for $0<\varepsilon<1$, a one-step positive process satisfying (H1)--(H2), with terminal value
\[
M_T=
\begin{cases}
\ell, & \text{with probability }1-\varepsilon,\\
R_\varepsilon, & \text{with probability }\varepsilon .
\end{cases}
\]
The terminal relation (H2), $\langle M_T\rangle=1$, fixes
\[
R_\varepsilon=
\frac{1-(1-\varepsilon)e^{-a}}{\varepsilon}.
\]
The corresponding entropy production is
\[
\Sigma_T=
\begin{cases}
a, & \text{with probability }1-\varepsilon,\\
-\ln R_\varepsilon, & \text{with probability }\varepsilon .
\end{cases}
\]
The positive maximum is bounded by $a$, while the mean entropy production is
\[
\langle \Sigma_T\rangle
=
(1-\varepsilon)a-\varepsilon\ln R_\varepsilon .
\]
As $\varepsilon\to0$, $R_\varepsilon\sim (1-e^{-a})/\varepsilon$, so $\varepsilon\ln R_\varepsilon\to0$ and therefore $\langle \Sigma_T\rangle\to a$. Thus $\sup\{\langle\Sigma_T\rangle:A_T\le a\}=a$.
This proves optimality for deterministic upper bounds that depend only on the ceiling $A_T\le a$, not optimality relative to richer information such as the full law of $A_T$.

This construction shows why one-sided information is insufficient. The process can produce entropy close to $a$ on almost every trajectory, while the fluctuation relation is maintained by extremely rare trajectories with very negative entropy production. The positive entropy-production ceiling does not control the exponential weight carried by the negative branch.

\subsection{Deterministic two-sided bound}
\label{sec:deterministic-two-sided}

A nontrivial upper bound appears once both sides of the entropy-production range are controlled. Suppose $-b\le \Sigma_T\le a$ for deterministic $a,b>0$. Equivalently, $e^{-a}\le M_T\le e^b$.
The fixed-time fluctuation theorem gives $\langle M_T\rangle=1$.

Let $f(m)=-\ln m$. On the interval $[e^{-a},e^b]$, the convex function $f$ lies below its secant line through the endpoint values
 $(e^{-a},a)$ and $(e^b,-b)$. Denote this deterministic secant line by $s_{a,b}(m)$, reserving $\ell_{L,H}(m)$ for secants written directly in the process's own running extrema $L,H$. Then $-\ln M_T\le s_{a,b}(M_T)$, and taking expectations with $\langle M_T\rangle=1$ gives
\[
\langle \Sigma_T\rangle
=
\langle -\ln M_T\rangle
\le
s_{a,b}(\langle M_T\rangle)
=
s_{a,b}(1).
\]
A direct calculation yields
\begin{equation}
\begin{split}
\langle \Sigma_T\rangle
\le
U(a,b)
&=
\frac{
a(e^b-1)-b(1-e^{-a})
}{
e^b-e^{-a}
} \\
&= a-(a+b)\frac{e^a-1}{e^{a+b}-1}.
\label{eq:deterministic_bound}
\end{split}
\end{equation}
The bound is sharp. Let
\[
M_T=
\begin{cases}
e^{-a}, & \text{with probability }p,\\
e^b, & \text{with probability }1-p .
\end{cases}
\]
The condition $\langle M_T\rangle=1$ gives
 $p=(e^b-1)/(e^b-e^{-a})$.
Then
\[
\Sigma_T=
\begin{cases}
a, & \text{with probability }p,\\
-b, & \text{with probability }1-p ,
\end{cases}
\]
and the mean entropy production equals $U(a,b)$. Pathwise, the same equality case can be realized by a martingale stopped when it first reaches either $e^{-a}$ or $e^b$.

The limit $b\to\infty$ recovers the one-sided result: $U(a,b)\to a$.
Thus the nontrivial part of the upper bound comes from controlling the negative entropy-production side. If arbitrarily negative entropy-production trajectories are allowed, the fluctuation relation can be satisfied by events of vanishing probability but arbitrarily large exponential weight.

The deterministic support bound in Eq.~\eqref{eq:deterministic_bound} constrains only the endpoint distribution: the condition $e^{-a}\le M_T\le e^b$ does not require any trajectory to realize both extrema along its path, and a martingale stopped at the two barriers, for example, hits only one of them on each path. The path-extrema bounds developed below instead track the realized interval $[L_T,H_T]$ for each trajectory, and so use additional path information.

\section{Random path-extrema upper bound}
\label{sec:random-envelope}

We now use the random interval $[L_T,H_T]$ realized by each path. First, convexity gives a bound conditioned on that interval. We then optimize over endpoint weights consistent with the retained extrema statistics and terminal normalization.

\subsection{Extrema-conditioned secant bound}

For fixed $(L_T,H_T)$, apply the secant inequality from Sec.~\ref{sec:setup} and take the conditional expectation. Using the endpoint weight $r_T$ from Eq.~\eqref{eq:r_actual} gives
\begin{equation}
\mathbb E[-\ln M_T\mid L_T,H_T]
\le
-\ln H_T
+
r_T\ln\frac{H_T}{L_T}.
\label{eq:conditional_secant_bound}
\end{equation}
Here the endpoint weight $r_T$ enters because the conditional mean of $M_T$ is $\mu_T=r_TL_T+(1-r_T)H_T$.

Finally, averaging Eq.~\eqref{eq:conditional_secant_bound} over the realized intervals gives
\begin{equation}
\langle \Sigma_T\rangle\le\Ucond,
\label{eq:Ucond_bound}
\end{equation}
where
\begin{equation}
\Ucond
=
\left\langle -\ln H_T\right\rangle
+
\left\langle
r_T\ln\frac{H_T}{L_T}
\right\rangle .
\label{eq:Ucond_def}
\end{equation}
Thus $\Ucond$ uses the actual conditional mean through $r_T$. If that information is discarded, only the total normalization cost in Eq.~\eqref{eq:rt_constraint} remains; we next allocate it across intervals.

In entropy variables, $L_T=e^{-A_T}$ and $H_T=e^{B_T}$, so $-\ln H_T=-B_T$ and $\ln(H_T/L_T)=A_T+B_T$.
Thus
\[
\Ucond
=
-\langle B_T\rangle
+
\langle r_T(A_T+B_T)\rangle .
\]
The weight $r_T$ measures the effective conditional bias toward the positive entropy-production endpoint $A_T$.

\subsection{Static upper bound and knapsack form}

Suppose only the joint law of $(L_T,H_T)$ is retained. To bound the mean without knowing $r_T$, optimize the secant payoff over all interval-dependent weights $r\in[0,1]$ satisfying the same normalization constraint. This static relaxation does not require the chosen weights to be realizable by a particular dynamics.

We define the path-extrema upper bound
\begin{multline}
\Uext
=
\left\langle -\ln H_T\right\rangle \\
+
\sup_{0\le r\le1}
\left\{
\left\langle
r\ln\frac{H_T}{L_T}
\right\rangle:
\left\langle r(H_T-L_T)\right\rangle
=
\langle H_T\rangle-1
\right\}.
\label{eq:u_ext_def}
\end{multline}
The actual $r_T$ is feasible, so $\langle \Sigma_T\rangle\le \Uext$.

The constraint in Eq.~\eqref{eq:u_ext_def} is always feasible. Since $L_T\le M_T\le H_T$ and $\langle M_T\rangle=1$,
 $\langle L_T\rangle\le 1\le \langle H_T\rangle$. Therefore $0\le \langle H_T\rangle-1\le \langle H_T-L_T\rangle$.
Hence there exists at least one interval-dependent weight function $r\in[0,1]$ satisfying the constraint. The actual conditional weight $r_T$ is one such function.

The variational problem in Eq.~\eqref{eq:u_ext_def} is linear. Let
\[
c_T=\ln\frac{H_T}{L_T},
\qquad
d_T=H_T-L_T,
\qquad
K=\langle H_T\rangle-1 .
\]
Then the optimization is
\[
\sup_{0\le r\le1}\langle r c_T\rangle
\]
subject to
\[
\langle r d_T\rangle=K .
\]
This is a continuous knapsack problem: moving terminal weight from $H_T$ to $L_T$ gains entropy $c_T$ at normalization cost $d_T$. The Supplemental Material derives the optimizer. Its efficiency ratio is
\begin{equation}
\rho_T=
\frac{\ln(H_T/L_T)}{H_T-L_T}.
\label{eq:rho_m}
\end{equation}

\begin{figure}
\includegraphics[width=\columnwidth]{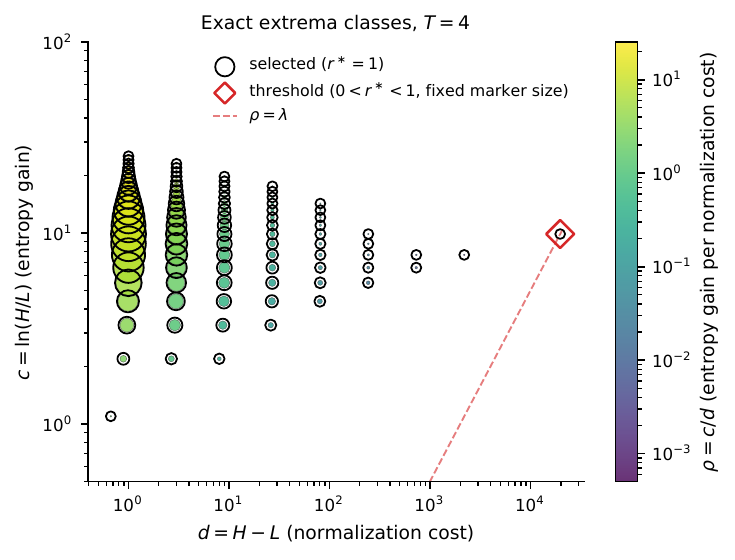}
\caption{Exact extrema classes for the three-state cycle at $T=4$ in cost-gain coordinates $d=H-L$, $c=\ln(H/L)$; color gives $\rho=c/d$. Marker area is proportional to class probability, including for selected classes (black circles). The fractional threshold class (red diamond) has a fixed marker size unrelated to its probability. Classes with probability at least $10^{-4}$ and the threshold class are shown; together they account for $99.87\%$ of the probability. The axes are cropped to contain all selected and threshold classes.}
\label{fig:knapsack}
\end{figure}

Figure~\ref{fig:knapsack} summarizes this construction by showing the realized extrema classes in cost-gain coordinates, the selected classes, and the threshold class in the fractional-knapsack rule.
In entropy variables, $\rho_T=(A_T+B_T)/(e^{B_T}-e^{-A_T})$.
The one-sided limits are
\[
\rho(A,0)=\frac{A}{1-e^{-A}},
\qquad
\rho(0,B)=\frac{B}{e^B-1},
\]
and $\rho(A,B)\to1$ as $A,B\to0$. For fixed $B$, $\rho(A,B)$ increases with $A$; for fixed $A$, it decreases with $B$.
The optimal weight function assigns $r=1$ first to extrema classes with the largest $\rho_T$, then to classes with progressively smaller $\rho_T$, with possible partial assignment at one threshold.

More explicitly, an optimal weight function $r_T^*$ has the threshold form
\[
r_T^*=
\begin{cases}
1, & \rho_T>\lambda,\\
0, & \rho_T<\lambda,\\
\theta_T\in[0,1], & \rho_T=\lambda,
\end{cases}
\]
where $\lambda$ and $\theta_T$ are chosen so that
\[
\left\langle r_T^*(H_T-L_T)\right\rangle=\langle H_T\rangle-1 .
\]
If a nonzero fraction of extrema classes has exactly the threshold value $\rho_T=\lambda$, the optimal weight function $r_T^*$ need not be unique. However, all choices that satisfy the same budget have the same objective value, so $\Uext$ and $\Agap$ are independent of the tie-breaking convention.
Equivalently, if $\rho^*(u)$ denotes the decreasing rearrangement of $\rho_T$ with respect to the measure $d\nu=(H_T-L_T)d\mathbb P$, meaning that extrema classes are listed from largest to smallest efficiency while weighted by their normalization cost,
then
\[
\Uext
=
\left\langle -\ln H_T\right\rangle
+
\int_0^{\langle H_T\rangle-1}\rho^*(u)\,du .
\]
This form emphasizes that the upper bound is determined by ranking extrema classes according to entropy span per normalization cost.

The path-extrema upper bound is a static relaxation: Eq.~\eqref{eq:u_ext_def} lets the endpoint assignment $r_T^*$ be chosen once the extrema class is known, rather than being produced by an actual path. Despite this, $\Uext$ is sharp already at the level of the minimal hypothesis (H1)--(H3). Since $M_0=1$, every (H1)--(H3) process has $L_T\le1\le H_T$ almost surely, and this sandwich is also sufficient: for any law of $(L,H)$ with $0<L\le1\le H<\infty$ a.s.\ and finite $\langle|\ln L|\rangle,\langle|\ln H|\rangle,\langle H\rangle$, there exists an (H1)--(H3) process with $(L_T,H_T)$ distributed as $(L,H)$ and $\langle\Sigma_T\rangle=\Uext$ exactly.

The construction chooses the terminal endpoint using the optimal weight $r^*(L,H)$, then visits both extrema on a piecewise-linear path. It satisfies the terminal normalization and attains $\Uext$; details are given in the Supplemental Material.

This construction uses no property of $M_t$ beyond (H1)--(H3): the path need be neither a martingale nor Markovian. Additional dynamical constraints -- martingality, or a specified Markov or diffusion dynamics -- can prevent it from being realized; that realizability problem is separate and is taken up in the Discussion. Section~\ref{sec:decomposition} shows that the resulting gap for any dynamically constrained process decomposes exactly into two nonnegative terms.

\section{Gap decomposition}
\label{sec:decomposition}

The difference between $\Uext$ and the actual mean separates into two nonnegative terms: the cost of a nonoptimal allocation across intervals and the cost of ending inside an interval.

Let $r_T^*$ denote an optimal weight function for the static problem in Eq.~\eqref{eq:u_ext_def}, that is, a choice of $r$ that attains the supremum. The actual process has the extrema-conditioned weight $r_T$ defined in Eq.~\eqref{eq:r_actual}.
Both $r_T$ and $r_T^*$ satisfy the same normalization-cost constraint, $\left\langle r(H_T-L_T)\right\rangle=\langle H_T\rangle-1$.

We define the cross-class allocation gap by
\begin{equation}
\Agap
=
\left\langle
(r_T^*-r_T)\ln\frac{H_T}{L_T}
\right\rangle .
\label{eq:A_gap}
\end{equation}
Because $r_T^*$ maximizes the linear functional in Eq.~\eqref{eq:u_ext_def} and $r_T$ satisfies the same constraint, we have $\Agap\ge 0$.
This term measures the gap from using the actual allocation $r_T$ rather than the class-optimal allocation $r_T^*$ across different realized path-extrema classes.

Next define the within-class curvature gap using the secant line $\ell_{L_T,H_T}$ defined in Sec.~\ref{sec:setup}:
\begin{equation}
\Cgap
=
\left\langle
\ell_{L_T,H_T}(M_T)-(-\ln M_T)
\right\rangle .
\label{eq:C_gap}
\end{equation}
Convexity of $-\ln m$ gives $\Cgap\ge0$.
This term measures the curvature loss from terminal values lying inside their realized intervals rather than at one of the two endpoints.

We now combine the definitions; the Supplemental Material spells out the algebra in full. By the definition of the static upper bound and the optimal weight function $r_T^*$,
\begin{equation}
\Uext
=
\left\langle -\ln H_T\right\rangle
+
\left\langle
r_T^*\ln\frac{H_T}{L_T}
\right\rangle .
\label{eq:Uext_opt_value}
\end{equation}
The actual conditional bound in Eq.~\eqref{eq:Ucond_def} is
\begin{equation}
\Ucond
=
\left\langle -\ln H_T\right\rangle
+
\left\langle
r_T\ln\frac{H_T}{L_T}
\right\rangle .
\label{eq:Ucond_actual_value}
\end{equation}
Subtracting Eq.~\eqref{eq:Ucond_actual_value} from Eq.~\eqref{eq:Uext_opt_value} and using the definition of $\Agap$ in Eq.~\eqref{eq:A_gap} gives
\begin{equation}
\Uext-\Ucond=\Agap .
\label{eq:Uext_Ucond_Agap}
\end{equation}
On the other hand, the extrema-conditioned secant construction gives
\begin{equation}
\left\langle
\ell_{L_T,H_T}(M_T)
\right\rangle
=
\Ucond,
\label{eq:secant_average_Ucond}
\end{equation}
so the definition of $\Cgap$ in Eq.~\eqref{eq:C_gap} gives
\begin{equation}
\Ucond-\langle \Sigma_T\rangle=\Cgap .
\label{eq:Ucond_mean_Cgap}
\end{equation}
Finally, combining Eqs.~\eqref{eq:Uext_Ucond_Agap} and \eqref{eq:Ucond_mean_Cgap} yields the central identity
\begin{equation}
\langle \Sigma_T\rangle
=
\Uext
-
\Agap
-
\Cgap .
\label{eq:main_decomposition}
\end{equation}

The intermediate bound $\Ucond$ makes the split meaningful: $\Agap$ measures the loss from the actual allocation across classes, and $\Cgap$ measures the secant loss within them. Thus the identity separates global allocation from local terminal placement.

The equality conditions are immediate. On the event $L_T=H_T$ the condition is satisfied trivially; on its complement, $\Cgap=0$ requires $M_T$ to equal one of the two endpoints almost surely. The allocation gap vanishes if and only if the actual extrema-conditioned weight $r_T$ is an optimal weight function for the static problem in Eq.~\eqref{eq:u_ext_def}. Thus the upper bound is attained exactly when terminal values lie at realized extrema and the endpoint allocation across extrema classes is extremal.

It is useful to make the curvature gap more quantitative. Since $f(m)=-\ln m$ satisfies $f''(m)=1/m^2$, for $m\in[L_T,H_T]$ we have
\[
\frac{1}{H_T^2}\le f''(m)\le \frac{1}{L_T^2}.
\]
The interpolation error for a twice differentiable convex function gives
\begin{multline*}
\frac{(H_T-m)(m-L_T)}{2H_T^2}
\le
\ell_{L_T,H_T}(m)-f(m) \\
\le
\frac{(H_T-m)(m-L_T)}{2L_T^2}.
\end{multline*}
Therefore
\[
\Cgap
\ge
\left\langle
\frac{(H_T-M_T)(M_T-L_T)}{2H_T^2}
\right\rangle .
\]
When the upper expectation is finite, one also has
\[
\Cgap
\le
\left\langle
\frac{(H_T-M_T)(M_T-L_T)}{2L_T^2}
\right\rangle .
\]
This lower bound shows explicitly that terminal values in the interior of their realized path-extrema intervals force a positive gap below the upper bound.

For a martingale or a specified Markov or diffusion dynamics, the path history can restrict the coupling of extrema and terminal value. Such restrictions can produce nonzero gaps even though the static bound is sharp in the unrestricted class.

\section{Examples}
\label{sec:examples}

A one-step positive martingale attains $\Uext$: its terminal value is an endpoint of its realized interval, and the actual endpoint allocation obeys the optimal efficiency threshold. A two-stage martingale tree can instead have both $\Agap>0$ and $\Cgap>0$. A separate static allocation example isolates $\Agap$ without claiming martingale realizability. These examples and their calculations are given in the Supplemental Material.

\subsection{Markov-jump realization with local detailed balance}
\label{sec:markov_jump_example}

We finally describe how the same quantities can be estimated in a genuine stochastic-thermodynamic model. Consider a continuous-time Markov jump process on three states arranged in a nonequilibrium cycle,
\[
1\rightleftarrows 2\rightleftarrows 3\rightleftarrows 1 .
\]
Let $k_{ij}$ denote the transition rate from state $i$ to state $j$. Local detailed balance assigns to each jump $i\to j$ the medium entropy increment
\[
\Delta s_{ij}=\ln\frac{k_{ij}}{k_{ji}} .
\]
For a trajectory $X_t$ observed in the nonequilibrium steady state $\pi$, the total entropy production over $[0,t]$ is
\[
\Sigma_t
=
\ln\frac{\pi(X_0)}{\pi(X_t)}
+
\sum_{0<\tau\le t}\ln\frac{k_{X_{\tau^-}X_{\tau}}}{k_{X_{\tau}X_{\tau^-}}} ,
\]
where the sum runs over jump times. In the steady state, $M_t=e^{-\Sigma_t}$ is the likelihood-ratio martingale between the forward trajectory measure and the time-reversed trajectory measure restricted to the observation interval. Therefore the path-extrema framework applies directly.

A minimal nonequilibrium steady-state example is obtained by taking clockwise rates
\[
k_{12}=k_{23}=k_{31}=k_+
\]
and counterclockwise rates
\[
k_{21}=k_{32}=k_{13}=k_- ,
\]
with $k_+>k_-$. The steady state is uniform by symmetry, so the system-entropy boundary term vanishes and each clockwise jump contributes $\ln(k_+/k_-)$ while each counterclockwise jump contributes $-\ln(k_+/k_-)$. Thus
\[
\Sigma_t
=
N_+(t)\ln\frac{k_+}{k_-}
-
N_-(t)\ln\frac{k_+}{k_-},
\]
where $N_+(t)$ and $N_-(t)$ are the numbers of clockwise and counterclockwise jumps.

\subsubsection{Exact finite-time benchmark}
\label{sec:exact_benchmark}

The net jump count $J_t=N_+(t)-N_-(t)$ is a biased random walk: jumps arrive at rate $k_++k_-$, and each is clockwise with probability $k_+/(k_++k_-)$. Since $\Sigma_t=J_t\ln(k_+/k_-)$ and $M_t=(k_+/k_-)^{-J_t}$, the extrema of $M_t$ follow from those of the walk. A dynamic program for the endpoint and running extrema, mixed over the Poisson jump count, provides the finite-time benchmark. It reproduces the analytic mean $\langle\Sigma_T\rangle=(k_+-k_-)T\ln(k_+/k_-)$, probability normalization, and the fluctuation relation to numerical precision. The Supplemental Material describes the truncated joint law and the enumeration, truncation-stability, and implementation checks.

\begin{figure}
\includegraphics[width=\columnwidth]{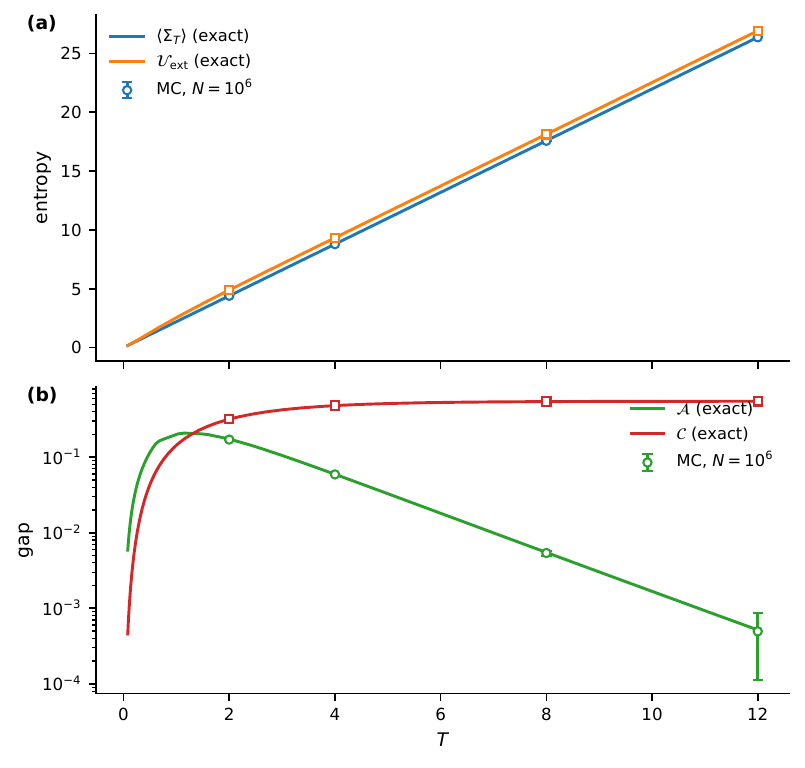}
\caption{Exact finite-time benchmark for the symmetric three-state cycle. Panel (a): exact $\langle\Sigma_T\rangle$ and $\Uext$ against $T$ (solid curves), with Monte Carlo estimates at $N=10^6$ total trajectories overlaid ($20$ batches; points with batch standard errors). Panel (b): exact allocation gap $\Agap$ and curvature gap $\Cgap$ (log scale), with the same Monte Carlo overlay. The Monte Carlo points in panel (a) lie close to the exact curves at $N=10^6$, but this proximity does not indicate that sampling has resolved the fluctuation relation. $\Uext$ depends on $\langle H_T\rangle$ through its normalization budget, and Fig.~\ref{fig:finite_sampling} shows that the empirical check $\langle e^{-\Sigma_T}\rangle_{\rm MC}$ remains far from its exact value $1$ at $T=8,12$ at this same sample size. Panel (b) shows growing relative uncertainty in the same Monte Carlo estimates at $T=8,12$.}
\label{fig:exact_benchmark}
\end{figure}

Table~\ref{tab:exact_benchmark} gives the exact benchmark for $k_+=3$, $k_-=1$ at the $T$ values used throughout this section. Figure~\ref{fig:exact_benchmark}(a) plots $\langle\Sigma_T\rangle$ and $\Uext$ against $T$, and panel (b) plots $\Agap$ and $\Cgap$. Over the range studied, the curvature gap grows and appears to level off, while the allocation gap rises and then falls by roughly two orders of magnitude between $T=2$ and $T=12$. At larger $T$ the allocation gap is small, although neither gap vanishes at any of the finite $T$ values checked.

\begin{table}[t]
\caption{Exact finite-time benchmark for the symmetric three-state cycle ($k_+=3$, $k_-=1$), from the dynamic program of the Supplemental Material. $\Agap$ and $\Cgap$ shown here are computed by the difference route ($\Uext-\Ucond$ and $\Ucond-\langle\Sigma_T\rangle$); a numerically separate direct-sum evaluation over the exact joint law agrees with these values to within $7\times10^{-15}$ at every $T$ shown (the Supplemental Material).}
\label{tab:exact_benchmark}
\begin{ruledtabular}
\begin{tabular}{c c c c c}
$T$ & $\langle\Sigma_T\rangle$ & $\Uext$ & $\Agap$ & $\Cgap$ \\
\hline
0.25 & 0.549306 & 0.600181 & 0.0415818 & 0.0092931 \\
0.5 & 1.09861 & 1.25752 & 0.115402 & 0.0435067 \\
1 & 2.19722 & 2.54026 & 0.199304 & 0.143728 \\
2 & 4.39445 & 4.88599 & 0.173482 & 0.318058 \\
4 & 8.78890 & 9.32904 & 0.0593717 & 0.480774 \\
8 & 17.5778 & 18.1266 & 0.00548371 & 0.543303 \\
12 & 26.3667 & 26.9160 & 0.000516711 & 0.548751 \\
\end{tabular}
\end{ruledtabular}
\end{table}

\subsubsection{Monte Carlo finite-sampling limitation}
\label{sec:finite_sampling}

The exact benchmark makes it possible to ask how well naive trajectory sampling would perform without it. We simulated ordinary Gillespie trajectories at $T\in\{2,4,8,12\}$, with $N\in\{10^3,10^4,10^5,10^6\}$ denoting the \emph{total} number of trajectories at each $(T,N)$ point, split into $20$ independent batches whose spread gives the standard errors. We formed the same plug-in estimators of $\langle\Sigma_T\rangle$, $\langle e^{-\Sigma_T}\rangle$, $\Uext$, $\Agap$, and $\Cgap$ that would be used without the exact solution, with no clipping or display sanitization: a negative empirical estimate of $\Agap$, when it occurs, is reported as negative.

\begin{figure}
\includegraphics[width=\columnwidth]{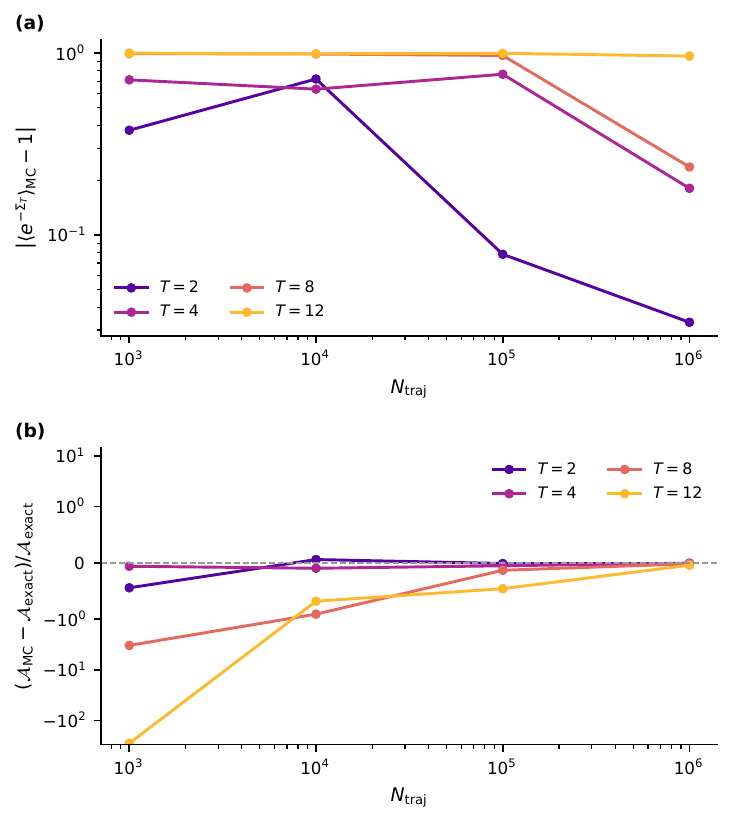}
\caption{Monte Carlo finite-sampling diagnostics against the exact benchmark. Panel (a): absolute deviation of the empirical fluctuation-relation check $\langle e^{-\Sigma_T}\rangle_{\rm MC}$ from its exact value of $1$, versus total trajectory count $N$ (split into $20$ batches, as in the main text), for $T=2,4,8,12$. Panel (b): signed relative error of the plug-in allocation-gap estimate $\Agap$ relative to its exact value, versus $N$ (symmetric log scale). No clipping is applied in either panel; where $\Agap^{\rm MC}$ is negative, the corresponding point in panel (b) is negative.}
\label{fig:finite_sampling}
\end{figure}

The fluctuation-relation check $\langle e^{-\Sigma_T}\rangle_{\rm MC}$ remains of order unity away from its exact value $1$ at $T=8$ and $T=12$ even at $N=10^6$ trajectories (Fig.~\ref{fig:finite_sampling}): the rare negative-entropy-production excursions that set $H_T$, and hence the normalization budget in $\Uext$, are not adequately sampled at long times. The plug-in estimate of $\Agap$ can land close to the exact value at large $N$ (Table~\ref{tab:exact_benchmark}, Fig.~\ref{fig:finite_sampling}b), but is erratic at small and moderate $N$: at $T=12$, $N=10^3$ it is negative and more than two orders of magnitude larger in size than the true gap. Its apparent convergence at $N=10^6$ therefore does not indicate that the fluctuation relation itself is resolved at that sample size. A small algebraic closure error is likewise not independent evidence of correct rare-event sampling, since all three decomposition terms are built from the same undersampled data. The exact solver instead cross-checks $\Agap$ and $\Cgap$ through numerically distinct direct sums, agreeing with the difference-based values to floating-point precision (the Supplemental Material).

This is the central empirical lesson of the example: at long times, the path-extrema construction is constrained by the same rare, entropy-consuming excursions that make the fluctuation relation itself hard to resolve empirically. Plug-in estimation of $\Uext$, $\Agap$, and $\Cgap$ from fully resolved trajectories is therefore not a shortcut around the sampling difficulty already present in verifying $\langle e^{-\Sigma_T}\rangle=1$; it inherits it. We return to the resulting practical interpretation of the bound in Sec.~\ref{sec:discussion}.

\subsubsection{Long-time behavior}

The mean entropy production grows linearly at rate
\[
\sigma
=3\pi_1 k_+\ln\frac{k_+}{k_-}
-3\pi_1 k_-\ln\frac{k_+}{k_-}
=(k_+-k_-)\ln\frac{k_+}{k_-},
\]
where $\pi_1=1/3$ in the symmetric steady state. Over the larger $T$ values in Table~\ref{tab:exact_benchmark}, $\Uext$ tracks $\langle\Sigma_T\rangle$ with a nearly constant offset, $\Cgap$ appears to level off, and $\Agap$ decreases. Whether these trends persist asymptotically remains open; neither gap is fixed by $\sigma$ alone.

\section{Discussion and conclusions}
\label{sec:discussion}

The random-extrema bound $\Uext$ extends the deterministic secant bound to realized path intervals, with a knapsack ranking that makes the normalization constraint explicit. Its slack separates exactly into the allocation and curvature gaps. The bound and decomposition require only (H1)--(H3); the construction in the Supplemental Material establishes sharpness in that unrestricted class.

Martingale constraints are stronger. A stopped martingale attains the deterministic two-sided support bound, and one-step martingales attain $\Uext$ for their own extrema laws. Neither example establishes attainment for an arbitrary prescribed extrema law: the path history and optional-stopping constraints restrict the coupling of extrema and terminal value. General martingale-constrained realizability remains open, as does realizability under specified Markov or diffusion dynamics; this is related to martingale transport \cite{BeiglbockJuillet2016}.

The exactly solved cycle illustrates the bound and shows why empirical use requires care: rare negative-entropy excursions can remain poorly sampled even when plug-in estimates appear close to their benchmark. The bound is a structural constraint, not a more efficient estimator of the mean from fully observed trajectories. Controlled rare-event methods and combinations with lower-bound inference are possible future directions. Whether the observed gap trends at long times are asymptotic remains open.

\bibliographystyle{apsrev4-2}
\bibliography{path_extrema_entropy_refs}
\end{document}


\title{Supplemental Material: Path-Extrema Upper Bounds on Mean Entropy Production}
\author{Surachate Limkumnerd}
\affiliation{Department of Physics, Faculty of Science, Chulalongkorn University, Bangkok 10330, Thailand}
\maketitle
Equation and section references to the main article use its numbering; references to appendices and examples below use the numbering of this supplement.
\section{Sharpness under the minimal hypothesis}

The construction is explicit. Let $r^*(L,H)$ be an optimal weight function for Eq.~\eqref{M-eq:u_ext_def}. Given a draw of $(L,H)$ and an independent uniform variable, set $M_T=L$ with probability $r^*(L,H)$ and $M_T=H$ otherwise. Take $M_t$ to be the piecewise-linear path from $M_0=1$ to whichever of $L,H$ is not the chosen $M_T$ over $[0,T/2]$, then from there to $M_T$ over $[T/2,T]$, collapsing a leg when its endpoints coincide (e.g.\ when $L=1$ or $H=1$). Since $L\le1\le H$, this path stays in $[L,H]$, is strictly positive, and has running extrema exactly $(L,H)$, so (H1) and (H3) hold with the prescribed law.

Saturation follows term by term. The budget constraint gives $\langle M_T\rangle=\langle r^*L+(1-r^*)H\rangle=\langle H\rangle-\langle r^*(H-L)\rangle=1$, so (H2) holds. Because $M_T\in\{L,H\}$ always, $\Cgap=0$; and because $\mu_T=r^*L+(1-r^*)H$ by construction, the actual weight $r_T=(H-\mu_T)/(H-L)$ equals $r^*$ exactly, so $\Agap=0$ as well. Hence $\langle\Sigma_T\rangle=\Uext-\Agap-\Cgap=\Uext$.

\section{Illustrative examples}
\subsection{Single-shot entropy fluctuation: saturation of the bound}

Consider a one-step positive martingale with $M_0=1$ and $M_T=Z$,
where $Z>0$ and $\langle Z\rangle=1$. This is the minimal fixed-time representation of an entropy-production fluctuation satisfying the integral fluctuation theorem:
 $\Sigma_T=-\ln Z$ and $\langle e^{-\Sigma_T}\rangle=\langle Z\rangle=1$.
This example represents a single-shot entropy-production measurement in which no intermediate time information is retained. It may be viewed as an effective description of a short experiment, a coarse-grained transition record, or a two-outcome entropy measurement. It is not yet a full continuous-time stochastic-thermodynamic model, because it keeps only the fixed-time entropy fluctuation and discards the dynamical path by which it was produced.

For this one-step process, $L_T=\min(1,Z)$ and $H_T=\max(1,Z)$.
In this degenerate one-step path, the only possible extrema are the initial value and the terminal value.
Thus $M_T$ always lies at one of the two endpoints of its realized interval:
 $M_T\in\{L_T,H_T\}$. Therefore the curvature gap vanishes, $\Cgap=0$.

The allocation gap also vanishes. If $Z<1$, then $M_T=L_T$ and $H_T=1$. The corresponding efficiency ratio is
\[
\rho_T=
\frac{\ln(1/Z)}{1-Z}>1 .
\]
If $Z>1$, then $M_T=H_T$ and $L_T=1$, and
\[
\rho_T=
\frac{\ln Z}{Z-1}<1 .
\]
Hence the actual endpoint assignment is a threshold rule in $\rho_T$: entropy-producing terminal outcomes occupy the high-$\rho_T$ classes, while entropy-consuming terminal outcomes occupy the low-$\rho_T$ classes. This is precisely the optimizer of the knapsack problem. Thus
 $\Agap=0$.
Equation~\eqref{M-eq:main_decomposition} gives
 $\langle \Sigma_T\rangle=\Uext$
for every one-step martingale.

This example is useful as an equality benchmark. When the entropy-production record has no internal temporal structure, the terminal value necessarily coincides with a realized path extremum, and the path-extrema upper bound is saturated.

\subsection{Mixture of operating modes: pure allocation gap}

The second example isolates the allocation gap $\Agap$. Consider an ensemble in which, before each run, a device is placed in one of two operating modes. Each mode has its own entropy-production window. Conditional on the mode, the terminal value lies at one of the two endpoints of that window, so there is no within-class curvature gap. The only possible gap is the allocation of entropy-producing endings across the two modes.

This example should be interpreted as a static extrema-allocation example, or support-window relaxation, rather than as a claim that a continuous-time martingale realizes exactly the specified joint law of $(L_T,H_T,M_T)$. Its role is to isolate $\Agap$ with $\Cgap=0$.

Let the two modes occur with probabilities $1/2$ and $1/2$. In mode 1, set
\[
L_1=\frac12,\qquad H_1=\frac32 .
\]
In mode 2, set
\[
L_2=\frac15,\qquad H_2=5 .
\]
Within each mode, let $M_T\in\{L_j,H_j\}$ with conditional probability $r_j$ of ending at $L_j$. We choose $r_j$ so that the conditional mean is one: $r_jL_j+(1-r_j)H_j=1$. Thus $r_j=(H_j-1)/(H_j-L_j)$. For the two modes, $r_1=\frac12$ and $r_2=\frac56$.
Each mode separately satisfies $\mathbb E[M_T\mid j]=1$, and therefore the full ensemble satisfies $\langle M_T\rangle=1$.

The mean entropy production is
\[
\langle \Sigma_T\rangle
=
\frac12\left[
\frac12\ln2-\frac12\ln\frac32
\right]
+
\frac12\left[
\frac56\ln5-\frac16\ln5
\right].
\]
Equivalently,
\[
\langle \Sigma_T\rangle
=
\frac14\ln\frac43+\frac13\ln5 .
\]
Numerically, $\langle \Sigma_T\rangle\simeq 0.6084$.

The efficiency ratios of the two mode windows are
\[
\rho_1=
\frac{\ln(H_1/L_1)}{H_1-L_1}
=
\ln3
\simeq 1.0986,
\]
and
\[
\rho_2=
\frac{\ln(H_2/L_2)}{H_2-L_2}
=
\frac{\ln25}{24/5}
\simeq 0.6706 .
\]
Thus mode 1 is the more efficient extrema class in the sense of Eq.~\eqref{M-eq:rho_m}. The path-extrema upper bound therefore assigns the entropy-producing endpoint first to mode 1.

The normalization budget is
\[
K=\langle H_T\rangle-1
=
\frac12\cdot\frac32+\frac12\cdot5-1
=
\frac94 .
\]
The cost of assigning all mode-1 trajectories to the lower endpoint is
\[
\frac12(H_1-L_1)=\frac12 .
\]
Thus the optimal allocation sets
\[
r_1^*=1 .
\]
The remaining cost is $9/4-1/2=7/4$. Since the full mode-2 cost is
\[
\frac12(H_2-L_2)=\frac{12}{5},
\]
the optimal allocation uses
\[
r_2^*=\frac{7/4}{12/5}=\frac{35}{48}.
\]
Therefore
\[
r_1^*=1,
\qquad
r_2^*=\frac{35}{48}.
\]

The resulting upper bound is
\[
\Uext
=
\left\langle -\ln H_T\right\rangle
+
\left\langle
r_T^*\ln\frac{H_T}{L_T}
\right\rangle,
\]
which gives numerically
\[
\Uext\simeq 0.7154 .
\]
Since terminal values are always at endpoints, $\Cgap=0$. Hence the entire gap is the allocation gap:
\[
\Agap
=
\Uext-\langle \Sigma_T\rangle
\simeq 0.1070 .
\]

This example shows that endpoint saturation alone does not guarantee saturation of the path-extrema upper bound. Even when $M_T$ always lies at $L_T$ or $H_T$, the ensemble may place entropy-producing terminal outcomes in extrema classes that are not optimal according to the efficiency ratio $\rho_T$. In physical terms, two operating modes may each obey the fluctuation theorem, but the full ensemble can still fall below the extremal bound because the entropy-producing outcomes are allocated inefficiently across modes.

\subsection{Two-stage martingale tree: allocation and curvature gaps}
\label{sec:two_stage_tree}

The third example is a genuine multistep martingale. It shows both a cross-class allocation gap and a within-class curvature gap; Appendix~\ref{app:tree} gives the full arithmetic.

Let
\[
M_0=1 .
\]
At the first step,
\[
M_1=
\begin{cases}
1/2, & \text{with probability }2/3,\\
2, & \text{with probability }1/3 .
\end{cases}
\]
The mean is one:
\[
\frac23\cdot\frac12+\frac13\cdot2=1 .
\]
At the second step, from $M_1=1/2$, set
\[
M_2=
\begin{cases}
1/4, & \text{with probability }1/2,\\
3/4, & \text{with probability }1/2 .
\end{cases}
\]
The conditional mean is $1/2$. From $M_1=2$, set
\[
M_2=
\begin{cases}
3/2, & \text{with probability }2/3,\\
3, & \text{with probability }1/3 .
\end{cases}
\]
The conditional mean is $2$. Thus $M_t$ is a positive martingale.

The four terminal paths are summarized in Table~\ref{tab:twostep_tree}.

\begin{table}[tbp]
\caption{Two-stage martingale tree. The terminal entropy production is $\Sigma_T=-\ln M_T$.}
\label{tab:twostep_tree}
\begin{ruledtabular}
\begin{tabular}{c c c c c}
Path & Probability & $M_T$ & $L_T$ & $H_T$ \\
\hline
$1\to 1/2\to 1/4$ & $1/3$ & $1/4$ & $1/4$ & $1$ \\
$1\to 1/2\to 3/4$ & $1/3$ & $3/4$ & $1/2$ & $1$ \\
$1\to 2\to 3/2$ & $2/9$ & $3/2$ & $1$ & $2$ \\
$1\to 2\to 3$ & $1/9$ & $3$ & $1$ & $3$
\end{tabular}
\end{ruledtabular}
\end{table}

For the four rows, the corresponding $(\Sigma_T,\rho_T)$ values are $(\ln4,4\ln4/3)$, $(\ln(4/3),2\ln2)$, $(-\ln(3/2),\ln2)$, and $(-\ln3,\ln3/2)$.

The actual mean entropy production is
\[
\langle \Sigma_T\rangle
=
\frac13\ln4
+
\frac13\ln\frac43
-
\frac29\ln\frac32
-
\frac19\ln3 .
\]
Numerically,
\[
\langle \Sigma_T\rangle\simeq 0.3458 .
\]

The efficiency ratios for the four realized intervals are
\[
\begin{array}{c|c}
(L_T,H_T) & \rho_T \\ \hline
(1/4,1) & \ln4/(3/4)\simeq 1.8484\\
(1/2,1) & \ln2/(1/2)\simeq 1.3863\\
(1,2) & \ln2\simeq 0.6931\\
(1,3) & \ln3/2\simeq 0.5493
\end{array}
\]
The path-extrema upper bound obtained from Eq.~\eqref{M-eq:u_ext_def} is
\[
\Uext\simeq 0.4363 .
\]
The gap is therefore
\[
\Uext-\langle \Sigma_T\rangle\simeq 0.0905 .
\]
Using Eqs.~\eqref{M-eq:A_gap} and \eqref{M-eq:C_gap}, one finds
\[
\Agap\simeq 0.0578,
\qquad
\Cgap\simeq 0.0327 .
\]
Thus
\[
0.0905\simeq 0.0578+0.0327,
\]
as required by Eq.~\eqref{M-eq:main_decomposition}.

The origin of $\Cgap$ is visible directly from the tree. Two terminal paths end strictly inside their realized intervals:
\[
1\to \frac12\to \frac34,
\qquad
1\to 2\to \frac32 .
\]
For the first of these,
\[
L_T=\frac12,\qquad M_T=\frac34,\qquad H_T=1.
\]
For the second,
\[
L_T=1,\qquad M_T=\frac32,\qquad H_T=2.
\]
On these paths, $L_T<M_T<H_T$, so the secant inequality is strict. These paths generate the within-class curvature gap.

The allocation gap $\Agap$ has a different origin. The relaxed upper bound assigns entropy-producing endpoints according to the global efficiency ranking of the extrema classes. The actual two-stage martingale cannot freely rearrange its terminal endpoints after the first branching event. Its endpoint allocation is constrained by its temporal structure, and this produces the cross-class allocation gap.

This example can be read as a minimal two-stage nonequilibrium experiment. The first stage creates an entropy-producing or entropy-consuming fluctuation. The second stage either amplifies that fluctuation or partially relaxes it back toward neutrality. The trajectories that relax toward the interior do not use the full entropy interval they have created, giving $\Cgap>0$. The branching structure also prevents the ensemble from allocating entropy-producing endings across extrema classes in the extremal way, giving $\Agap>0$.

\appendix

\section{Notation summary}
\label{app:notation}

For reference, we collect the main quantities used in the paper:
\[
M_t=e^{-\Sigma_t},
\qquad
L_T=\inf_{0\le t\le T}M_t,
\qquad
H_T=\sup_{0\le t\le T}M_t .
\]
The entropy-production extrema, conditional terminal mean, and actual extrema-conditioned endpoint weight are
\[
\begin{gathered}
A_T=-\ln L_T,\qquad
B_T=\ln H_T,\\
\mu_T=\mathbb E[M_T\mid L_T,H_T],\\
r_T=\frac{H_T-\mu_T}{H_T-L_T}.
\end{gathered}
\]
The path-extrema upper bound is
\begin{multline*}
\Uext
=
\left\langle -\ln H_T\right\rangle \\
+
\sup_{0\le r\le1}
\left\{
\left\langle
r\ln\frac{H_T}{L_T}
\right\rangle:
\left\langle r(H_T-L_T)\right\rangle
=
\langle H_T\rangle-1
\right\}.
\end{multline*}
The efficiency ratio is
\[
\rho_T=
\frac{\ln(H_T/L_T)}{H_T-L_T}
=
\frac{A_T+B_T}{e^{B_T}-e^{-A_T}}.
\]
The allocation and curvature gaps are
\[
\Agap
=
\left\langle
(r_T^*-r_T)\ln\frac{H_T}{L_T}
\right\rangle,
\]
and
\[
\Cgap
=
\left\langle
\ell_{L_T,H_T}(M_T)-(-\ln M_T)
\right\rangle .
\]
The central decomposition is
\[
\langle \Sigma_T\rangle
=
\Uext
-
\Agap
-
\Cgap .
\]

\section{Continuous-knapsack optimizer}
\label{app:knapsack}

This appendix gives the standard derivation of the optimizer used in Eq.~\eqref{M-eq:u_ext_def}. Let
\[
c_T=\ln\frac{H_T}{L_T},
\qquad
d_T=H_T-L_T,
\qquad
K=\langle H_T\rangle-1 .
\]
The optimization problem is
\[
\sup_{0\le r\le1}\langle r c_T\rangle
\]
subject to
\[
\langle r d_T\rangle=K .
\]
Assume first that $d_T>0$ almost surely on the nontrivial part of the sample space. Define
\[
\rho_T=\frac{c_T}{d_T}.
\]
Then
\[
\langle r c_T\rangle
=
\langle r\rho_T d_T\rangle .
\]
Introduce the measure
\[
d\nu=d_T\,d\mathbb P .
\]
The constraint and objective become
\[
\int r\,d\nu=K,
\qquad
\int r\rho_T\,d\nu .
\]
Thus the optimizer chooses $r=1$ on the largest values of $\rho_T$, then on progressively smaller values, with possible partial assignment at the threshold. Equivalently, if $\rho^*(u)$ denotes the decreasing rearrangement of $\rho_T$ with respect to $\nu$, then
\[
\sup_{0\le r\le1,\ \int r\,d\nu=K}
\int r\rho_T\,d\nu
=
\int_0^K \rho^*(u)\,du .
\]
This gives the rearrangement form
\[
\Uext
=
\left\langle -\ln H_T\right\rangle
+
\int_0^{\langle H_T\rangle-1}\rho^*(u)\,du .
\]

The threshold form follows equivalently from a Lagrange multiplier. Consider
\[
\langle r c_T\rangle-\lambda\langle r d_T\rangle
=
\left\langle r(c_T-\lambda d_T)\right\rangle .
\]
Pointwise maximization over $0\le r\le1$ gives
\[
r_T^*=
\begin{cases}
1, & c_T-\lambda d_T>0,\\
0, & c_T-\lambda d_T<0,\\
\theta_T\in[0,1], & c_T-\lambda d_T=0.
\end{cases}
\]
Since $c_T-\lambda d_T=d_T(\rho_T-\lambda)$, this is the threshold rule stated in the main text.

\section{Algebra of the gap decomposition}
\label{app:decomposition}

Here we spell out the algebra behind Eq.~\eqref{M-eq:main_decomposition}. Let
\[
\mu_T=\mathbb E[M_T\mid L_T,H_T]
\]
and
\[
r_T=\frac{H_T-\mu_T}{H_T-L_T}.
\]
The secant line of $f(m)=-\ln m$ over $[L_T,H_T]$ is
\[
\ell_{L_T,H_T}(m)
=
\frac{H_T-m}{H_T-L_T}f(L_T)
+
\frac{m-L_T}{H_T-L_T}f(H_T).
\]
Conditioning on $(L_T,H_T)$ gives
\begin{multline*}
\mathbb E[\ell_{L_T,H_T}(M_T)\mid L_T,H_T] \\
=
\frac{H_T-\mu_T}{H_T-L_T}f(L_T)
+
\frac{\mu_T-L_T}{H_T-L_T}f(H_T).
\end{multline*}
Using the definition of $r_T$,
\begin{equation*}
\begin{split}
	\mathbb E[\ell_{L_T,H_T}(M_T)\mid L_T,H_T]
	&= r_T(-\ln L_T) \\
	& \qquad\qquad +(1-r_T)(-\ln H_T) \\
	&= -\ln H_T+ r_T\ln\frac{H_T}{L_T}.
\end{split}	
\end{equation*}
Averaging gives
\[
\left\langle \ell_{L_T,H_T}(M_T)\right\rangle
=
\Ucond.
\]
By definition,
\[
\Cgap=
\left\langle
\ell_{L_T,H_T}(M_T)-(-\ln M_T)
\right\rangle,
\]
so $\langle \Sigma_T\rangle=\Ucond-\Cgap$.

Similarly,
\[
\Uext
=
\left\langle -\ln H_T\right\rangle
+
\left\langle
r_T^*\ln\frac{H_T}{L_T}
\right\rangle,
\]
while
\[
\Ucond
=
\left\langle -\ln H_T\right\rangle
+
\left\langle
r_T\ln\frac{H_T}{L_T}
\right\rangle .
\]
Thus
\[
\Uext-\Ucond
=
\left\langle
(r_T^*-r_T)\ln\frac{H_T}{L_T}
\right\rangle
=
\Agap .
\]
Combining these identities yields
\[
\langle \Sigma_T\rangle
=
\Uext-\Agap-\Cgap .
\]

\section{Details for the two-stage tree}
\label{app:tree}

For the two-stage martingale tree in Sec.~\ref{sec:two_stage_tree}, the four terminal outcomes are
\begin{multline*}
(M_T,L_T,H_T,p)
=
\left(\frac14,\frac14,1,\frac13\right),
\left(\frac34,\frac12,1,\frac13\right), \\
\left(\frac32,1,2,\frac29\right),
\left(3,1,3,\frac19\right).
\end{multline*}
The actual mean entropy production is
\[
\langle \Sigma_T\rangle
=
\frac13\ln4
+
\frac13\ln\frac43
-
\frac29\ln\frac32
-
\frac19\ln3
\simeq 0.345821 .
\]

The normalization budget is $K=\langle H_T\rangle-1$. Here
\[
\langle H_T\rangle
=
\frac13(1)+\frac13(1)+\frac29(2)+\frac19(3)
=
\frac{13}{9},
\]
so $K=4/9$.
The extrema classes, costs, and efficiency ratios are
\[
\begin{array}{c|c|c|c}
(L,H) & p & p(H-L) & \rho=\ln(H/L)/(H-L)\\
\hline
(1/4,1) & 1/3 & 1/4 & \ln4/(3/4)\\
(1/2,1) & 1/3 & 1/6 & \ln2/(1/2)\\
(1,2) & 2/9 & 2/9 & \ln2\\
(1,3) & 1/9 & 2/9 & \ln3/2
\end{array}
\]
The costs of the first two classes add to $1/4+1/6=5/12$. The remaining budget is $4/9-5/12=1/36$. Thus the optimal allocation uses the first two classes fully and uses a fraction $(1/36)/(2/9)=1/8$ of the third class. Hence $r^*=(1,1,1/8,0)$ on the four classes in decreasing order of $\rho$.

The path-extrema upper bound is therefore
\[
\Uext
=
\left\langle -\ln H_T\right\rangle
+
\left\langle r^*\ln\frac{H_T}{L_T}\right\rangle .
\]
The first term is $\left\langle -\ln H_T\right\rangle=-\frac29\ln2-\frac19\ln3$, and the second term is $\frac13\ln4+\frac13\ln2+\frac29\cdot\frac18\ln2$. Therefore
\[
\Uext
=
\frac13\ln4
+
\left(\frac13+\frac1{36}-\frac29\right)\ln2
-
\frac19\ln3 .
\]
Numerically,
\[
\Uext\simeq 0.436301 .
\]

We next compute the actual conditional allocation weights $r_T$. In this discrete example, each extrema class contains one terminal outcome, so
\[
r_T=\frac{H_T-M_T}{H_T-L_T}.
\]
Thus $r_T=(1,1/2,1/2,0)$ on the four classes. The allocation gap is
\[
\Agap
=
\left\langle
(r^*-r_T)\ln\frac{H_T}{L_T}
\right\rangle .
\]
Only the second and third classes contribute:
\[
\Agap
=
\frac13\left(1-\frac12\right)\ln2
+
\frac29\left(\frac18-\frac12\right)\ln2 .
\]
Thus $\Agap=(1/6-1/12)\ln2=(1/12)\ln2\simeq 0.057762$.

The curvature gap is
\[
\Cgap
=
\left\langle
\ell_{L_T,H_T}(M_T)-(-\ln M_T)
\right\rangle .
\]
Only the two interior terminal outcomes contribute. For $(L,H,M)=(1/2,1,3/4)$,
\[
\ell_{1/2,1}(3/4)
=
\frac12\ln2,
\]
while
\[
-\ln(3/4)=\ln\frac43 .
\]
The contribution is $\frac13\left[\frac12\ln2-\ln\frac43\right]$.
For $(L,H,M)=(1,2,3/2)$,
\[
\ell_{1,2}(3/2)
=
-\frac12\ln2,
\]
while
\[
-\ln(3/2)=-\ln\frac32 .
\]
The contribution is $\frac29\left[-\frac12\ln2+\ln\frac32\right]$.
Hence
\[
\Cgap
=
\frac13\left[
\frac12\ln2-\ln\frac43
\right]
+
\frac29\left[
-\frac12\ln2+\ln\frac32
\right]
\simeq 0.032718 .
\]
Therefore
\[
\Agap+\Cgap
\simeq
0.090480
=
\Uext-\langle \Sigma_T\rangle,
\]
as stated in the main text.

\section{Exact finite-time solution for the three-state cycle}
\label{app:exact_cycle}

This appendix gives the exact finite-time solution used in Sec.~\ref{M-sec:exact_benchmark}, for the symmetric cycle with clockwise rate $k_+$ and counterclockwise rate $k_-$ at every state.

\subsection{Reduction to a biased random walk}

Because every state has total escape rate $k_++k_-$, the jump times of the cycle form a Poisson process of rate $k_++k_-$, independent of the jump directions, and each jump is independently clockwise with probability $p=k_+/(k_++k_-)$ or counterclockwise with probability $q=1-p$, regardless of the current state. Writing $N$ for the total number of jumps up to time $T$, $N\sim\mathrm{Poisson}((k_++k_-)T)$, and conditional on $N=n$, the number of clockwise jumps is $\mathrm{Binomial}(n,p)$. The net winding number $J_t=N_+(t)-N_-(t)$ is therefore, conditional on $N=n$, a simple random walk of $n$ steps with up-probability $p$, and $\Sigma_t=J_t\ln(k_+/k_-)$, $M_t=e^{-\Sigma_t}=(k_+/k_-)^{-J_t}$. Since $M_t$ is monotonically decreasing in $J_t$, the running extrema of $M_t$ are determined by the running extrema of $J_t$: with $m_t=\min_{0\le s\le t}J_s$ and $M^{\max}_t=\max_{0\le s\le t}J_s$,
\[
L_T=(k_+/k_-)^{-M^{\max}_T},
\qquad
H_T=(k_+/k_-)^{-m_T}.
\]

\subsection{Dynamic program over the joint law of the walk and its running extrema}

Fix $n$ and let $(j,m,M)$ denote the current position and running minimum/maximum of the embedded walk after some number of steps $\le n$. Each further step updates
\[
(j,m,M)\to(j+1,m,\max(M,j+1))
\]
with probability $p$, and
\[
(j,m,M)\to(j-1,\min(m,j-1),M)
\]
with probability $q$, starting from $(0,0,0)$. We compute the joint law of $(j,m,M)$ after $n=0,1,\dots,n_{\max}$ steps by a forward sweep, implemented as a bounded dense array indexed by $(j,m,M)$ with vectorized shifts for the two update rules. At each step $n$, the current joint law is mixed into running totals weighted by $\mathrm{Poisson}(n;(k_++k_-)T)$, for every $T$ of interest simultaneously, so a single sweep to $n_{\max}$ produces the benchmark at all $T$ values used in the manuscript.

The two decomposition terms that ultimately matter, $\Uext$ and $\Cgap$, depend on the joint law of $(j,m,M)$ only through the marginal law of $(m,M)$ and the conditional mean $\mu(m,M)=\mathbb E[(k_+/k_-)^{-j}\mid m,M]$: because the secant line $\ell_{L,H}$ used to define $\Cgap$ is affine in its argument, $\mathbb E[\ell_{L,H}(M_T)\mid L_T,H_T]=\ell_{L,H}(\mu(m,M))$ exactly, with no residual conditional-variance term. It therefore suffices to accumulate, at each step $n$ and for each $T$, the two $(m,M)$-resolved arrays $P(m,M)=\Pr[\text{running min}=m,\text{running max}=M]$ and $S(m,M)=\mathbb E[(k_+/k_-)^{-j}\mathbb 1(m,M)]$ (so $\mu=S/P$), together with the scalar $\mathbb E[j]$ used as an independent numerical check on $\langle\Sigma_T\rangle$. From these, $\Uext$, $\Ucond$, $\Agap=\Uext-\Ucond$, and $\Cgap=\Ucond-\langle\Sigma_T\rangle$ are computed exactly as in Sec.~\ref{M-sec:decomposition}, using the same fractional-knapsack routine as the Monte Carlo estimator.

\subsection{Numerical tolerance and self-checks}

The horizon $n_{\max}$ is chosen so that the Poisson upper tail $\Pr[N>n_{\max}]$ is below $10^{-13}$, with headroom added; for $T=12$, $k_+=3$, $k_-=1$, this gives $n_{\max}=140$. This bounds the omitted probability mass, not by itself the error in extrema-weighted observables. We performed the following checks. (i) For small $n_{\max}$ (up to $n=7$), the dynamic program agrees element-by-element with brute-force enumeration of all $2^n$ jump sequences to floating-point precision. (ii) At every $T$ used in the manuscript, both the computed probability mass and the fluctuation-relation check equal $1$ to within $10^{-10}$ or better, and the computed $\langle\Sigma_T\rangle$ agrees with $(k_+-k_-)T\ln(k_+/k_-)$ to within $3\times10^{-13}$. (iii) Increasing $n_{\max}$ by a factor of $1.5$ changes every reported quantity ($\langle\Sigma_T\rangle$, $\Uext$, $\Agap$, $\Cgap$) by at most $7\times10^{-15}$ at $T=12$. This is an observed stability check, not a rigorous bound on the remaining truncation error. (iv) We also computed the gaps by numerically distinct routes: differences $\Agap=\Uext-\Ucond$ and $\Cgap=\Ucond-\langle\Sigma_T\rangle$, and a direct sum $\sum_i P_i(r_i^*-r_i)c_i$ for $\Agap$, together with a bin-resolved entropy accumulator for $\Cgap$. The routes agree within $7\times10^{-15}$ at every $T$ checked. They use the same underlying joint law and are mathematically equivalent, so this agreement checks the implementation and floating-point evaluation rather than independently validating the theorem or the joint law.